\documentstyle[prl,aps,preprint]{revtex}

\begin{document}
\draft
\widetext

\title{Gravitational Duality in MacDowell-Mansouri Gauge Theory}
\author{H. Garc\'{\i}a-Compe\'an$^{a}$\thanks{Present Address:
{\it School of Natural Sciences,
Institute for Advanced Study, Olden Lane, Princeton NJ 08540 USA}. E-mail:
compean@sns.ias.edu}, O. Obreg\'on$^{b}$\thanks{E-mail:
octavio@ifug3.ugto.mx} and C. Ram\'{\i}rez$^c$\thanks{E-mail:
cramirez@fcfm.buap.mx}\\
$^{a}$ {\it Departamento de F\'{\i}sica, Centro de Investigaci\'on y de
Estudios
Avanzados del IPN\\
 P.O. Box 14-740, 07000, M\'exico D.F., M\'exico}\\
$^b$ {\it Instituto de F\'{\i}sica de la Universidad de Guanajuato\\
 P.O. Box E-143, 37150, Le\'on Gto., M\'exico}\\
$^c$ {\it  Facultad de Ciencias F\'{\i}sico Matem\'aticas\\
Universidad Aut\'onoma de Puebla\\
P.O. Box 1364, 72000, Puebla, M\'exico}}

\maketitle

\begin{abstract}
Strong-weak duality invariance can only be defined for particular sectors
of supersymmetric Yang-Mills theories.  Nevertheless, for full non-Abelian
non-supersymmetric theories, dual theories with inverted couplings, have
been found.  We show that an analogous procedure allows to find the dual
action to the gauge theory of gravity constructed by the
MacDowell-Mansouri model plus the superposition of a $\Theta$ term.
\end{abstract}

%\pacs{PACS numbers: } %\hspace{7cm}

\narrowtext
\newpage

%\section{Introduction}

In the search for the construction of an unified model for fundamental
interactions different routes have been followed in the past.  A natural
one is to consider higher dimensional models of gravity (and supergravity)
\cite{review} as candidates for a unified model of the fundamental
interactions.  The metric, in the higher dimensional space-time, is
considered the basic field by means of which one can construct the fields
that describe the fundamental interactions and gravity itself in the four
dimensional world.

In contrast, if one searches for a unified framework of the non-gravitational fundamental
interactions in four space-time dimensions, the usual way is that the elementary
interactions are described in terms of a connection associated with an
internal group leaving space-time nondynamical.  However, the
evolution of the metric of this space-time should describe precisely
gravity.  Various attempts have been made to construct Yang-Mills type
gauge theories of gravity \cite{west}, where the {\it basic} fields are
the gauge fields of an appropriate group $G$.  The metric (the tetrad) and
the Lorentz connection are obtained only as components of these fields, so
that standard general relativity is a consequence of the proposed gauge
theory.  Following this scheme twenty years ago, MacDowell and Mansouri
(MM) succeeded in constructing a gauge theory of gravity \cite{Mac}.
Pagels \cite{pagels} gave an Euclidean formulation.  Inspired  in the MM work, another approach has been constructed \cite{nieto}, based only on a
self-dual spin connection (where duality is defined with respect to the
corresponding group indices).  This proposal generalizes and includes the
MM gauge theory of gravity and the Pleba\'nski-Ashtekar formulation
\cite{jacobson}.

On the other hand,  it is well known that for non-Abelian non-supersymmetric
gauge theories, ``$S$-dual" theories can be constructed which results in
a kind of ``massive" non-linear sigma models \cite{freedman}.  The starting
 Yang-Mills theory contains a $CP$-violating $\theta$-term and results to
be equivalent to the linear combination of the actions corresponding to
the self-dual and antiself-dual field strengths
\cite{witten,mohammedi,ganor,lozano}.

Knowing the self-dual formulation of the MM theory \cite{nieto}, it is tempting to
search if by combining it linearly with the antiself-dual part one gets
the MM action plus a $\Theta$-term and then try to proceed as in
Yang-Mills theories to obtain a dual model.  It turns out that all this
works, as we will show in this paper.

Let us review the MM proposal.  The starting point in the construction of
this theory is to consider an SO(3,2) gauge theory with Lie algebra-valued
gauge potential $A^{AB}_\mu$ where the indices $\mu = 0, 1, 2, 3$ are
space-time indices and the indices $A, B= 0, 1, 2, 3, 4$.

>From the gauge potential $A^{~AB}_\mu$ we may introduce the corresponding
field strength

\begin{equation}
F_{\mu\nu}^{~~AB} = \partial_\mu A_\nu^{~AB} - \partial_\nu A^{AB}_\mu
+ \frac{1}{2} f^{AB}_{CDEF} A^{CD}_\mu A^{EF}_\nu,
\end{equation}
where $f^{AB}_{CDEF}$ are the structure constants of SO(3,2).  MM
choose $F^{a4}_{\mu\nu} \equiv 0$ and as an action

\begin{equation}
S_{MM} = \int d^4 x \epsilon^{\mu\nu\alpha\beta} \epsilon_{abcd}
F^{ab}_{\mu\nu} F^{cd}_{\alpha\beta}
\end{equation}
where $a,b,...{\rm etc.}=0,1,2,3.$

On the other hand, a generalization now taking the self-dual (or
antiself-dual) part of the connection has been proposed \cite{nieto} and
by this means the corresponding field strengths ${^+}F^{ab}_{\mu\nu}$ (or
${^-}F^{ab}_{\mu\nu})$ have been constructed.

Moreover, one knows that for Yang-Mills theories, a kind of ``dual
theories" with inverted couplings can be found
\cite{mohammedi,ganor,lozano}.  One starts with the Yang-Mills action plus
a $CP$ violating $\theta$-term.  It can be shown that these two terms are
equivalent to a linear combination of the self-dual and antiself-dual
Yang-Mills actions.

One can then search whether the construction of a linear combination of
the corresponding self-dual and antiself-dual parts of the
MacDowell-Mansouri action can also be reduced to the standard MM action
plus a kind of $\Theta$-term and, moreover, if by this means one can find
the ``dual-theory" associated with the MM theory.

Let us consider the action

\begin{equation}
S=\int d^4 x \epsilon^{\mu\nu\alpha\beta} \epsilon_{abcd} \bigg(
{^+} \tau {^+}F^{ab}_{\mu\nu} {^+}F^{cd}_{\alpha\beta} - {^-} \tau
{^-}F^{ab}_{\mu\nu} {^-}F^{cd}_{\alpha\beta} \bigg) ,
\end{equation}
where

\begin{equation}
{^\pm}F^{~~ab}_{\mu\nu} =  \frac{1}{2} \bigg( F^{~~ab}_{\mu\nu}
\pm \tilde F^{ad}_{\mu\nu} \bigg),
\end{equation}
with $\tilde F^{ab}_{\mu\nu} = - \frac{1}{2} i \epsilon^{ab}_{~~cd}
F^{cd}_{\mu\nu}$.
It can easily be shown \cite{samuel},
that this action can be rewritten as

\begin{equation}
S=\frac{1}{2} \int d^{4} x \epsilon^{\mu\nu\alpha\beta} \epsilon_{abcd}
\bigg[ ({^+}\tau - {^-}\tau) F^{ab}_{\mu\nu} F^{cd}_{\alpha\beta} + (
{^+}\tau + {^-}\tau) F^{ab}_{\mu\nu} \tilde F^{cd}_{\alpha\beta} \bigg].
\end{equation}
MM showed in their original paper \cite{Mac} that the first term in this
action reduces to the Euler topological term plus the Einstein-Hilbert
action with a cosmological  term.  This was achieved after identifying
the components of the gauge field $A^{~AB}_{\mu}$ with the Ricci rotation
coefficients and the vierbein.   Similarly the second term can be
shown  to be equal to $iP$ where $P$ is the Pontrjagin topological
term \cite{nieto}.  Thus, this term is a genuine $\Theta$ term  with $\Theta$
given by the sum ${^+}\tau + {^-}\tau$.

Our second task is to find the ``dual theory", in a similar sense as
in Yang-Mills theories \cite{mohammedi,ganor}.  For that purpose
we consider the parent action

\begin{equation}
I= \int d^{4}x \epsilon^{\mu\nu\alpha\beta} \epsilon_{abcd}
\bigg( c_1{^+} G^{ab}_{\mu\nu} {^+}G^{cd}_{\alpha\beta} + c_2
{^-} G^{ab}_{\mu\nu} {^-} G^{cd}_{\alpha\beta} + c_3 {^+}
F^{ab}_{\mu\nu} {^+}G^{cd}_{\alpha\beta} + c_4 {^-} F^{ab}_{\mu\nu}
{^-} G^{cd}_{\alpha\beta} \bigg) .
\end{equation}
>From which obviously as in Yang-Mills theories the action (3) can be
recovered after integration on ${^+}G$ and ${^-}G$.

In order to get the ``dual theory" we follow reference \cite{ganor}.
  Therefore,  one should start with the partition function

\begin{equation}
Z= \int {\cal D} {^+}G \, {\cal D} {^-}G \, {\cal D} A\,\,e^{-I}.
\end{equation}
To proceed with  the integration over the gauge fields we observe that

\begin{equation}
F^{ab}_{\mu\nu} = \partial_\mu A^{ab}_{\nu} - \partial_\nu A^{ab}_\mu
+ \frac{1}{2} f^{ab}_{CDEF} A^{CD}_{\mu} A^{EF}_{\nu} .
\end{equation}
Taking into account the explicit expressions for the structure constants
\cite{Mac} the second term of the right hand side will naturally split in
four  terms given by

\begin{equation}
A^{~ad}_{\mu} A^{~~~b}
_{\nu d} - A^{~ad}_\nu A^{~~~b}_{\mu d} - \lambda^2 \bigg( A^{~a4}_\mu
A^{~b4}_\nu - A^{~a4}_\nu A^{~b4}_\mu \bigg).
\end{equation}
Furthermore,  given the definition of ${^\pm}F$ in Eq. (4) and
the fact that ${^\pm} G$ are (anti) self-dual fields one can rewrite
the last two terms in Eq. (6) in the following manner

\begin{equation}
4i \epsilon^{\mu\nu\alpha\beta} F^{~~ab}_{\mu\nu} \bigg(
c_3 {^+}G_{\alpha\beta ab} - c_4 {^-}G_{\alpha\beta ab} \bigg).
\end{equation}
Now we are ready to perform the integration over the gauge field
$A$ in the partition function (7).  The integration over the
components $A^{~a4}_\mu$ is given by a  Gaussian integral which turns out
to be $det {\bf G}^{-1/2}$, where  {\bf G} is a matrix given by

\begin{equation}
{\bf G}^{\mu\nu}_{ab} = 8 i \lambda^2 \epsilon^{\mu\nu\alpha\beta}
\bigg( c_3 {^+}G_{\alpha\beta ab} - c_4 {^-}G_{\alpha\beta ab} \bigg).
\end{equation}

Thus, the partition function (7) can be written as

\begin{equation}
Z= \int {\cal D}{^+} G\, {\cal D} {^-}G \, {\cal D} A^{ab}_\mu\,\, det {\bf
G}^{-1/2}\, e^{- \, {I\!\!I}}
\end{equation}

where

$${I\!\!I}=2 i \int d^4 x  \epsilon^{\mu\nu\alpha\beta} \bigg[ c_1
{^+}G^{ab}_{\mu\nu} {^+}G_{\alpha\beta ab} - c_2 {^-} G^{ab}_{\mu\nu}
{^-}G_{\alpha\beta ab} + 2 H^{~~ab}_{\mu\nu} ( c_3
{^+} G_{\alpha\beta ab} - c_4 {^-}G_{\alpha\beta ab}) \bigg], $$
with

\begin{equation}
H^{ab}_{\mu\nu} = \partial_\mu A^{ab}_\nu - \partial_\nu A^{ab}_\mu
+ \frac{1}{2} f^{ab}_{cdef} A^{cd}_\mu A^{ef}_\nu
\end{equation}
is the SO(3,1) field strength \cite{Mac}.

Our last step to get the dual action is the integration over
$A^{ab}_\mu$.  This kind of integration is well known and has been
performed in previous works \cite{mohammedi,ganor}.  The result is

\begin{equation}
Z=\int {\cal D}{^+}G\, {\cal D} {^-}G \,\, det {\bf G}^{-1/2} \,\,
det ({^+}M)^{-1/2} det ({^-}M)^{-1/2} \,\, e^{- \int d^4x \tilde L}
\end{equation}
with

\begin{equation}
\begin{array}{ll}
\tilde{L} &= \epsilon^{\mu\nu\rho\sigma} \bigg[ -{1 \over 4
{^+}\tau} {^+}G^{ab}_{\mu \nu} {^+}G_{\rho\sigma ab} + {1 \over 4
{^-}\tau} {^-}G^{ab}_{\mu \nu} {^-}G_{\rho \sigma ab} +
2 \partial_{\nu} {^+}G_{\rho \sigma ab} {({^+}M)}^{-1 \ abcd}_{\mu
\lambda} \epsilon^{\lambda \theta \alpha \beta}
\partial_{\theta} {^+}G_{\alpha \beta cd} \\
&- 2 \partial_{\nu} {^-}G_{\rho
\sigma
ab} {({^-}M)}^{-1 \ abcd}_{\mu
\lambda} \epsilon^{\lambda \theta \alpha \beta}
\partial_{\theta} G^-_{\alpha \beta cd}\bigg]
\end{array}
\label{tcinco}
\end{equation}
and

$${^\pm}M^{\mu\nu cd}_{ab} = \frac{1}{2} \epsilon^{\mu\nu}_{~~\alpha\beta}
\bigg( - \delta^c_a {^\pm}G^{~\alpha\beta d}_{b} + \delta^c_b
{^\pm}G^{~\alpha\beta d}_a + \delta^d_a {^\pm} G^{~\alpha\beta c}_{b}
- \delta^d_b {^\pm} G^{~\alpha\beta c}_a \bigg), $$
with ${^+} \tau = - \frac{1}{4c_1},$ \    ${^-}\tau = - \frac{1}{4c_2}$,
$c_3 = c_4 = 1.$

Now we come back to our formulae (8), (9) and (13).  We immediately observe
that

\begin{equation}
F^{ab}_{\mu\nu} = H^{ab}_{\mu\nu} - \lambda^2 \Sigma^{ab}_{\mu\nu},
\end{equation}
with $\Sigma^{ab}_{\mu\nu} = A^{a4}_{\mu} A^{b4}_\nu - A_\nu^{a4}
A^{b4}_{\mu}$.
As mentioned MM were able to reduce their theory to standard general
relativity with cosmological constant and the Euler topological term
by identifying $A_\mu^{~ab}$ with the spin connection $ \omega^{~ab}_\mu$ and $A_\mu^{~a4}$
with the tetrad $e^{~a}_\mu$. Obviously then $H^{~~ab}_{\mu\nu} \equiv R^{ab}_{\mu\nu}$.

In a previous work we took a non-dynamical topological action
constructed by the sum of the Euler and the Pontrjagin topological
terms \cite{garcia}. The resulting partition function,  constructed with the
dual Lagrangian, in that case results formally equal to Eq. (14) but
without
the $det {\bf G}^{-1/2}$.  On the one hand,  this new term results
from the extended number of the gauge field components, the $A^{a4}_\mu$.
It also manifests extra conditions over the integration on the fields
${^+}G$ and ${^-}G$ in the partition function (14).  At the level of the
equations of motion this condition manifests itself as

\begin{equation}
\epsilon_{\mu\nu\rho\tau }{\bf G}^{~~\mu\nu\rho}_{\alpha} = 0,
\end{equation}
relating the components of the $G$ field.  This condition does not appear
in the previous topological model \cite{garcia}.  The
non-dynamical model considered in the previous work  \cite{garcia}
results in a kind of non-linear sigma model \cite{mohammedi,ganor} of the
type considered by Freedman and Townsend \cite{freedman,lozano},
as the usual Yang-Mills dual models.  The dual
to the dynamical gravitational model (15) considered here results in a
Lagrangian of the same structure.  However, it differs from these previous
cases by the features discussed above.

It is worth to mention that considering as starting point the O(5) group,
Pagels \cite{pagels}, by means of a gauge fixing, has got a
theory similar to that of MacDowell-Mansouri.  Essentially the same arguments
followed in this work hold for Pagel's approach.

We feel that in the framework of the already shown ``gravitational
duality" interesting questions arise:

 {\bf 1)} The structure that we have shown is as already argued, similar
to that in Yang-Mills theories.  This points out to the possible existence
of ``gravitational monopoles" and/or solitons \cite{alvarez}.

{\bf 2)} It is of interest to
consider supergravity versions of the work presented here. In general supersymmetry improves the mathematical consistence, and improves also the duality properties of Yang-Mills theories \cite{seiberg}.
 The self-dual extension of the MM $N=1$
supergravity has been already obtained \cite{nieto2}.  So the
supersymmetric equivalent to Eq. (3) can be straightforwardly constructed
and then its ``$S$-dual" can be searched for.

{\bf 3)} It will be also of interest to investigate the relation of the
``gravitational duality" of this paper with the gravitational duality
proposed by Hull by means of the new gravitational branes arising in type
II superstrings and M theory \cite{hull}.

The previous points are under current investigation and  will be
reported elsewhere.

%\vskip 2truecm

%\centerline{\bf Acknowledgments}
%This work has been partially  supported by CONACyT
%
%\vskip 2truecm
%

%\end{document}
\end{document}